\documentstyle[prd,aps,preprint,tighten,epsfig]{revtex}

\begin{document}

\draft

\title{Lepton-number-violating decays of singly-charged Higgs
bosons \\
in the minimal type-(I+II) seesaw model at the TeV scale}
\author{{\bf Ping Ren} \thanks{E-mail: renp@ihep.ac.cn}
and {\bf Zhi-zhong Xing}
\thanks{E-mail: xingzz@ihep.ac.cn}}
\address{Institute of High Energy Physics and Theoretical Physics
Center for Science Facilities, \\
Chinese Academy of Sciences, P.O. Box 918, Beijing 100049, China}

\maketitle

\begin{abstract}
The lepton-number-violating decays of singly-charged Higgs bosons
$H^{\pm}$ are investigated in the minimal type-(I+II) seesaw model
with one $SU(2)_L$ Higgs triplet $\Delta$ and one heavy Majorana
neutrino $N^{}_1$ at the TeV scale. We find that the branching
ratios ${\cal B} (H^{+} \to l^+_{\alpha}{\bar \nu}^{}_{})$ (for
$\alpha = e, \mu, \tau$) depend not only on the mass and mixing
parameters of three light neutrinos $\nu^{}_i$ (for $i=1,2,3$) but
also on those of $N^{}_1$. Assuming the mass of $N^{}_1$ to lie in
the range of 200 GeV to 1 TeV, we figure out the generous
interference bands for the contributions of $\nu^{}_i$ and
$N^{}_1$ to ${\cal B} (H^{+} \to l^+_{\alpha}{\bar \nu}^{}_{})$.
We illustrate some salient features of such interference effects
by considering three typical mass patterns of $\nu^{}_i$, and show
that the relevant Majorana CP-violating phases can affect the
magnitudes of ${\cal B} (H^{+} \to l^+_{\alpha}{\bar \nu}^{}_{})$
in this parameter region.
\end{abstract}


\section{Introduction}

As the Large Hadron Collider (LHC) will soon bring us to a new
energy frontier, major discoveries of new physics beyond the
Standard Model (SM) at the TeV scale are highly anticipated
\cite{TEV}. Indeed, the observation of solar and atmospheric
neutrino oscillations has provided us with the first convincing
evidence for new physics beyond the SM \cite{PDG}; i.e., three known
neutrinos are massive and their flavors mix with one another.
Whether the origin of non-zero but tiny neutrino masses can be
understood at the LHC is an open but interesting question. It has
recently been conjectured that possible new physics, if it exists at
the TeV scale and is responsible for the electroweak symmetry
breaking, might also be relevant to the neutrino mass generation
\cite{ICHEP08}.

The conventional seesaw picture \cite{Seesaw1}, named nowadays as
the type-I seesaw mechanism, gives a natural explanation of the
smallness of neutrino masses by introducing a few heavy
right-handed Majorana neutrinos. Another popular way to generate
tiny neutrino masses, the so-called type-II seesaw mechanism, is
to extend the SM by including one $SU(2)_L$ Higgs triplet
\cite{Seesaw2}. One may also combine the two scenarios by assuming
the existence of both the Higgs triplet and right-handed Majorana
neutrinos, leading to a more general seesaw mechanism which has
several different names in the literature \cite{Venice}. To avoid
any literal confusion, here we follow some authors and simply
refer to this ``hybrid" seesaw scenario as the type-(I+II) seesaw
mechanism. The gauge-invariant neutrino mass terms in a
type-(I+II) seesaw model can be written as
\begin{eqnarray}
-{\cal L}^{}_{\rm mass} \; = \; \overline{l^{}_{\rm L}} Y^{}_\nu
\tilde{H} N^{}_{\rm R} + \frac{1}{2} \overline{N^{c}_{\rm R}}
M^{}_{\rm R} N^{}_{\rm R} + \frac{1}{2} \overline{l^{}_{\rm L}}
Y^{}_\Delta \Delta i\sigma^{}_2 l^c_{\rm L} + {\rm h.c.} \; ,
\end{eqnarray}
where $M^{}_{\rm R}$ is the mass matrix of right-handed Majorana
neutrinos, and
\begin{equation}
\Delta \; \equiv \; \left(\matrix{H^- & - \sqrt{2} ~ H^0 \cr
\sqrt{2} ~ H^{--} & -H^-}\right)
\end{equation}
denotes the $SU(2)_L$ Higgs triplet. After the spontaneous gauge
symmetry breaking, we obtain the neutrino mass matrices $M^{}_{\rm
D} = Y^{}_\nu v/\sqrt{2}$ and $M^{}_{\rm L} = Y^{}_\Delta
v^{}_\Delta$, where $\langle H \rangle \equiv v/\sqrt{2}$ and
$\langle \Delta \rangle \equiv v^{}_\Delta$ correspond to the
vacuum expectation values of the neutral components of $H$ and
$\Delta$. To minimize the degrees of freedom associated with
$M^{}_{\rm L}$, $M^{}_{\rm D}$ and $M^{}_{\rm R}$, one may assume
that there is only one heavy Majorana neutrino (denoted as
$N^{}_1$) in the model with $M^{}_{\rm R}$ and $M^{}_{\rm D}$
being $1\times 1$ and $3\times 1$ respectively. Such a simplified
seesaw scenario is phenomenologically viable
\cite{Gu,Chan,Ren,Zhou} and can be referred to as the {\it
minimal} type-(I+II) seesaw model, whose simplicity makes it
interesting and instructive to reveal some salient features of the
type-(I+II) seesaw mechanism. We shall focus our attention on this
simple case in the present paper.

Our purpose is to investigate the lepton-number-violating decays of
singly-charged Higgs bosons $H^{\pm}$ in the minimal type-(I+II)
seesaw model. Such decays can naturally happen because $\Delta$ is
allowed to couple to the standard-model Higgs doublet $H$ and thus
the lepton number is violated by two units \cite{Seesaw2}. If the
mass scale of $\Delta$ is of ${\cal O}(1)$ TeV or smaller, then both
$H^{\pm\pm}$ and $H^{\pm}$ can be produced at the LHC via the
Drell-Yan process $q\bar{q} \rightarrow \gamma^*, Z^* \rightarrow
H^{++} H^{--}$ and through the charged-current process
$q\bar{q}^\prime \rightarrow W^* \rightarrow H^{\pm\pm}H^{\mp}$. In
some optimistic scenarios, one can investigate different seesaw
models by searching for the clean signals of lepton number violation
in the decays of doubly- and singly-charged Higgs bosons at the TeV
scale \cite{Ren,Zhou,KS,Fileviez,Triplet}. When it comes to large
$Y^{}_\Delta$ and small $v^{}_{\Delta}$ (say, $v_\Delta < 10^{-4}$
GeV), the dominant decay channels of $\Delta$ will be the leptonic
modes \cite{Fileviez}, such as $H^{++} \rightarrow l_{\alpha}^+
l_{\beta}^+$ and $H^{+} \rightarrow l_{\alpha}^+ \bar{\nu}$ (for
$\alpha, \beta = e, \mu, \tau$). An analysis of $H^{\pm\pm}
\rightarrow l_{\alpha}^\pm l_{\beta}^\pm$ decays in the minimal
type-(I+II) seesaw model has been done in Ref. \cite{Ren}. Here we
are going to calculate the branching ratios of $H^{+} \rightarrow
l_{\alpha}^+ \bar{\nu}$ and $H^{-} \rightarrow l_{\alpha}^- \nu$ in
the same model. The importance of the lepton-number-violating decays
of $H^\pm$ has been emphasized in Ref. \cite{Fileviez} within the
type-II seesaw framework. Our interest is to explore the interplay
between type-I and type-II seesaw terms in $H^{+} \rightarrow
l_{\alpha}^+ \bar{\nu}$ or $H^{-} \rightarrow l_{\alpha}^- \nu$
decays within the type-(I+II) seesaw framework.

Following Ref. \cite{Fileviez}, we obtain the decay rates of $H^{+}
\to l^+_{\alpha}{\bar \nu}^{}_{\beta}$ as
\begin{equation}
\Gamma(H^{+} \to l^+_{\alpha}{\bar \nu}^{}_{\beta}) \; = \;
\frac{1}{4\pi} ~ |\left(Y^{}_{\Delta}\right)^{}_{\alpha\beta}|^2
M^{}_{H^{+}} \; .
\end{equation}
The branching ratios of $H^{+} \to l^+_{\alpha}{\bar
\nu}^{}_{\beta}$ turn out to be \cite{Fileviez}
\begin{equation}
{\cal B}(H^{+} \to l^+_{\alpha} {\bar \nu}^{}_{}) \equiv
\sum^{}_{\beta} {\cal B}( H^{+} \to l^+_{\alpha} {\bar
\nu}^{}_{\beta} ) \equiv \frac{\displaystyle \sum^{}_{\beta}
\Gamma(H^{+} \to l^+_{\alpha} {\bar \nu}^{}_{\beta})} {\displaystyle
\sum^{}_{\rho,\sigma}\Gamma(H^{+} \to l^+_\rho {\bar
\nu}^{}_\sigma)} = \frac{\displaystyle \sum^{}_{\beta}
|\left(M^{}_{\rm L}\right)^{}_{\alpha\beta}|^2} {\displaystyle
\sum^{}_{\rho,\sigma}|\left(M^{}_{\rm L}\right)^{}_{\rho\sigma}|^2}
\; ,
\end{equation}
where the Greek subscripts run over $e$, $\mu$ and $\tau$. It
becomes obvious that the magnitudes of ${\cal B}(H^{+} \to
l^+_{\alpha} {\bar \nu}^{}_{})$ are only relevant to the matrix
elements of $M^{}_{\rm L}$. Note that the matrix elements of
$M^{}_{\rm L}$ rely both on the mass and mixing parameters of three
light neutrinos $\nu^{}_i$ (for $i=1,2,3$) and on those of $N^{}_1$
in the minimal type-(I+II) seesaw model \cite{Ren}. When the
contribution of $N^{}_1$ to ${\cal B}(H^{+} \to l^+_{\alpha} {\bar
\nu}^{}_{})$ is negligibly small, our result can simply reproduce
that obtained in the type-II seesaw model \cite{Fileviez}. But when
type-I and type-II seesaw terms are comparable in magnitude, we have
to take care of their significant interference effects. Assuming the
mass of $N^{}_1$ to lie in the range of 200 GeV to 1 TeV, we figure
out the generous interference bands for the contributions of
$\nu^{}_i$ and $N^{}_1$ to ${\cal B} (H^{+} \to l^+_{\alpha}{\bar
\nu}^{}_{})$. We illustrate some salient features of such
interference effects by considering three typical mass patterns of
$\nu^{}_i$. We also show that the relevant Majorana CP-violating
phases can affect the magnitudes of ${\cal B} (H^{+} \to
l^+_{\alpha}{\bar \nu}^{}_{})$, unlike the case in the type-II
seesaw mechanism \cite{Fileviez}. Although our numerical results are
subject to the minimal type-(I+II) seesaw model, they can serve as a
good example to illustrate the interplay between light and heavy
Majorana neutrinos in a generic type-(I+II) seesaw scenario.

\section{Interference bands and Majorana phases}

After the spontaneous electroweak symmetry breaking, we rewrite Eq.
(1) as
\begin{eqnarray}
-{\cal L}^\prime_{\rm mass} \; = \; \frac{1}{2} ~ \overline{\left(
\nu^{}_{\rm L} ~N^c_{\rm R}\right)} ~ \left( \matrix{ M^{}_{\rm L} &
M^{}_{\rm D} \cr M^T_{\rm D} & M^{}_{\rm R}}\right) \left( \matrix{
\nu^c_{\rm L} \cr N^{}_{\rm R}}\right) + {\rm h.c.} \; .
\end{eqnarray}
We assume the existence of only a single heavy Majorana neutrino
$N^{}_1$. The $4\times 4$ neutrino mass matrix in Eq. (5) is
symmetric and can be diagonalized by the following unitary
transformation:
\begin{eqnarray}
\left(\matrix{V & R \cr S & U}\right)^\dagger \left( \matrix{
M^{}_{\rm L} & M^{}_{\rm D} \cr M^T_{\rm D} & M^{}_{\rm R}}\right)
\left(\matrix{V & R \cr S & U}\right)^*  = \left( \matrix{
\widehat{M}^{}_\nu & {\bf 0} \cr {\bf 0} & M^{}_1}\right) \; ,
\end{eqnarray}
where $\widehat{M}^{}_\nu = {\rm Diag}\{m^{}_1, m^{}_2, m^{}_3\}$
with $m^{}_i$ being the masses of three light neutrinos $\nu^{}_i$
and $M^{}_1$ denotes the mass of $N^{}_1$. Following Ref.
\cite{Xing08}, we parametrize $V$ and $R$ as
\begin{eqnarray}
V & = & \left( \matrix{ c^{}_{14} & 0 & 0 \cr -\hat{s}^{}_{14}
\hat{s}^*_{24} & c^{}_{24} & 0 \cr - \hat{s}^{}_{14} c^{}_{24}
\hat{s}^*_{34} & - \hat{s}^{}_{24} \hat{s}^*_{34} & c^{}_{34} \cr}
\right) \left( \matrix{ c^{}_{12} c^{}_{13} & \hat{s}^*_{12}
c^{}_{13} & \hat{s}^*_{13} \cr -\hat{s}^{}_{12} c^{}_{23} -
c^{}_{12} \hat{s}^{}_{13} \hat{s}^*_{23} & c^{}_{12} c^{}_{23} -
\hat{s}^*_{12} \hat{s}^{}_{13} \hat{s}^*_{23} & c^{}_{13}
\hat{s}^*_{23} \cr \hat{s}^{}_{12} \hat{s}^{}_{23} - c^{}_{12}
\hat{s}^{}_{13} c^{}_{23} & -c^{}_{12} \hat{s}^{}_{23} -
\hat{s}^*_{12} \hat{s}^{}_{13} c^{}_{23} & c^{}_{13} c^{}_{23}
\cr} \right) \; , \nonumber \\
R & = & \left( \matrix{ \hat{s}^*_{14} \cr c^{}_{14} \hat{s}^*_{24}
\cr c^{}_{14} c^{}_{24} \hat{s}^*_{34} \cr} \right) \; ,
\end{eqnarray}
where $c^{}_{ij} \equiv \theta^{}_{ij}$, $s^{}_{ij} \equiv
\sin\theta^{}_{ij}$ and $\hat{s}^{}_{ij} \equiv
e^{i\delta^{}_{ij}} s^{}_{ij}$ with $\theta^{}_{ij}$ and
$\delta^{}_{ij}$ (for $1\leq i < j \leq 4$) being the rotation
angles and phase angles, respectively. If the heavy Majorana
neutrino $N^{}_1$ is decoupled (i.e., $\theta^{}_{14} =
\theta^{}_{24} = \theta^{}_{34} = 0$), $V$ will become a unitary
matrix and take the standard form \cite{PDG}. Hence non-vanishing
$R$ measures the non-unitarity of $V$.

Now we make use of Eqs. (6) and (7) to reconstruct the matrix
elements of $M^{}_{\rm L}$ in terms of $m^{}_i$, $M^{}_1$, $V$ and
$R$. It is easy to obtain $M^{}_{\rm L} \; = \; V \widehat{M}^{}_\nu
V^T + M^{}_1 R R^T $. Taking the approximation $c^{}_{13} \approx
c^{}_{i4} \approx 1$ based on current experimental constraints
$s^{}_{13} < 0.16$ \cite{Fogli} and $s^{}_{i4} \lesssim 0.1$ (for
$i=1,2,3$) \cite{Antusch}, we arrive at
\begin{eqnarray}
\sum_\beta |\left(M^{}_{\rm L}\right)^{}_{e\beta}|^2 & = & m^2_1
c^2_{12} + m^2_2 s^2_{12} + m^2_3 s^2_{13} + M^{}_1
s^2_{14} \left( s^2_{14} + s^2_{24} + s^2_{34} \right) \nonumber \\
&& + 2 m^{}_1 M^{}_1 {\rm Re} \left[ c^{}_{12} \hat{s}^{}_{14}
\left( c^{}_{12} \hat{s}^{}_{14} - \hat{s}^{}_{12} c^{}_{23}
\hat{s}^{}_{24} + \hat{s}^{}_{12} \hat{s}^{}_{23} \hat{s}^{}_{34}
\right) \right] \nonumber \\
&& + 2 m^{}_2 M^{}_1 {\rm Re} \left[ \hat{s}^{*}_{12}
\hat{s}^{}_{14} \left( \hat{s}^{*}_{12} \hat{s}^{}_{14} + c^{}_{12}
c^{}_{23} \hat{s}^{}_{24} - c^{}_{12}
\hat{s}^{}_{23}\hat{s}^{}_{34} \right) \right] \nonumber \\
&& + 2 m^{}_3 M^{}_1 {\rm Re} \left[ \hat{s}^{*}_{13}
\hat{s}^{}_{14} \left( \hat{s}^{*}_{13} \hat{s}^{}_{14} +
\hat{s}^{*}_{23} \hat{s}^{}_{24} + c^{}_{23} \hat{s}^{}_{34} \right)
\right] \; , \nonumber \\
\sum_\beta |\left(M^{}_{\rm L}\right)^{}_{\mu\beta}|^2 & = & m^2_1
s^2_{12} c^2_{23} + m^2_2 c^2_{12} c^2_{23} + m^2_3 s^2_{23} +
M^{}_1 s^2_{24} \left( s^2_{14} + s^2_{24} + s^2_{34} \right) \nonumber \\
&& - 2 m^{}_1 M^{}_1 {\rm Re} \left[ \hat{s}^{}_{12} c^{}_{23}
\hat{s}^{}_{24} \left( c^{}_{12} \hat{s}^{}_{14} - \hat{s}^{}_{12}
c^{}_{23} \hat{s}^{}_{24} + \hat{s}^{}_{12} \hat{s}^{}_{23}
\hat{s}^{}_{34} \right) \right] \nonumber \\
&& + 2 m^{}_2 M^{}_1 {\rm Re} \left[ c^{}_{12} c^{}_{23}
\hat{s}^{}_{24} \left( \hat{s}^{*}_{12} \hat{s}^{}_{14} + c^{}_{12}
c^{}_{23} \hat{s}^{}_{24} - c^{}_{12} \hat{s}^{}_{23}
\hat{s}^{}_{34} \right) \right] \nonumber \\
&& + 2 m^{}_3 M^{}_1 {\rm Re} \left[ \hat{s}^{*}_{23}
\hat{s}^{}_{24} \left( \hat{s}^{*}_{13} \hat{s}^{}_{14} +
\hat{s}^{*}_{23} \hat{s}^{}_{24} + c^{}_{23} \hat{s}^{}_{34} \right)
\right] \; , \nonumber \\
\sum_\beta |\left(M^{}_{\rm L}\right)^{}_{\tau\beta}|^2 & = & m^2_1
s^2_{12} s^2_{23} + m^2_2 c^2_{12} s^2_{23} + m^2_3 c^2_{23} +
M^{}_1 s^2_{34} \left( s^2_{14} + s^2_{24} + s^2_{34} \right) \nonumber \\
&& + 2 m^{}_1 M^{}_1 {\rm Re} \left[ \hat{s}^{}_{12} \hat{s}^{}_{23}
\hat{s}^{}_{34} \left( c^{}_{12} \hat{s}^{}_{14} - \hat{s}^{}_{12}
c^{}_{23} \hat{s}^{}_{24} + \hat{s}^{}_{12}
\hat{s}^{}_{23} \hat{s}^{}_{34} \right) \right] \nonumber \\
&& - 2 m^{}_2 M^{}_1 {\rm Re} \left[ c^{}_{12} \hat{s}^{}_{23}
\hat{s}^{}_{34} \left( \hat{s}^{*}_{12} \hat{s}^{}_{14} + c^{}_{12}
c^{}_{23} \hat{s}^{}_{24} - c^{}_{12} \hat{s}^{}_{23}
\hat{s}^{}_{34} \right) \right] \nonumber \\
&& + 2 m^{}_3 M^{}_1 {\rm Re} \left[ c^{}_{23} \hat{s}^{}_{34}
\left( \hat{s}^{*}_{13} \hat{s}^{}_{14} + \hat{s}^{*}_{23}
\hat{s}^{}_{24} + c^{}_{23} \hat{s}^{}_{34} \right) \right] \; ;
\end{eqnarray}
and
\begin{eqnarray}
\sum_{\rho,\sigma}|\left(M^{}_{\rm L}\right)^{}_{\rho\sigma}|^2 & =
& \left(m^2_1 + m^2_2 + m^2_3\right) + M^2_1 \left( s^2_{14} +
s^2_{24} + s^2_{34} \right)^2 \nonumber \\
&& + 2 m^{}_1 M^{}_1 {\rm Re} \left[ \left( c^{}_{12}
\hat{s}^{}_{14} - \hat{s}^{}_{12} c^{}_{23} \hat{s}^{}_{24} +
\hat{s}^{}_{12} \hat{s}^{}_{23} \hat{s}^{}_{34} \right)^2 \right]
\nonumber \\
&& + 2 m^{}_2 M^{}_1 {\rm Re} \left[ \left( \hat{s}^{*}_{12}
\hat{s}^{}_{14} + c^{}_{12} c^{}_{23} \hat{s}^{}_{24} - c^{}_{12}
\hat{s}^{}_{23} \hat{s}^{}_{34} \right)^2 \right] \nonumber \\
&& + 2 m^{}_3 M^{}_1 {\rm Re} \left[ \left( \hat{s}^{*}_{13}
\hat{s}^{}_{14} + \hat{s}^{*}_{23} \hat{s}^{}_{24} + c^{}_{23}
\hat{s}^{}_{34} \right)^2 \right] \; .
\end{eqnarray}
By combining Eqs. (8) and (9) with Eq. (4), we are then able to
calculate the branching ratios ${\cal B} (H^{+} \to
l^+_{\alpha}{\bar \nu}^{}_{})$. Note that these branching ratios can
also be expressed in terms of the branching ratios ${\cal B}(H^{++}
\rightarrow l^+_\alpha l^+_\beta)$ obtained in Ref. \cite{Ren};
namely,
\begin{eqnarray}
{\cal B}(H^+ \to e^+ \bar{\nu}) & = & {\cal B}(H^{++} \to e^+ e^+) +
\frac{1}{2} \left[ {\cal B}(H^{++} \to e^+ \mu^+) + {\cal B}(H^{++}
\to e^+ \tau^+) \right] \; , \nonumber \\
{\cal B}(H^+ \to \mu^+ \bar{\nu}) & = & {\cal B}(H^{++} \to \mu^+
\mu^+) + \frac{1}{2} \left[ {\cal B}(H^{++} \to e^+ \mu^+) + {\cal
B}(H^{++} \to \mu^+ \tau^+) \right] \; , \nonumber \\
{\cal B}(H^+ \to \tau^+ \bar{\nu}) & = & {\cal B}(H^{++} \to \tau^+
\tau^+) + \frac{1}{2} \left[ {\cal B}(H^{++} \to e^+ \tau^+) + {\cal
B}(H^{++} \to \mu^+ \tau^+) \right] \; .
\end{eqnarray}
If the heavy Majorana neutrino $N^{}_1$ is essentially decoupled
(i.e., $\theta^{}_{i4} \approx 0$ for $i=1,2,3$), then the unitarity
of $V$ will be restored. In this case, the results of ${\cal B}
(H^{+} \to l^+_{\alpha}{\bar \nu}^{}_{})$ are the same as those
given in the type-II seesaw model \cite{Fileviez}.

If the contributions of $\nu^{}_i$ and $N^{}_1$ to $(M^{}_{\rm
L})^{}_{\alpha\beta}$ are comparable in magnitude, there will be
significant interference effects on the branching ratios of $H^{+}
\to l^+_{\alpha}{\bar \nu}^{}_{}$ decays. To be explicit, we take
$\Delta m^2_{21} \sim 7.7 \times 10^{-5} ~ {\rm eV}^2$ and $|\Delta
m^2_{32}| \sim 2.4 \times 10^{-3} ~ {\rm eV}^2$ \cite{Fogli} as the
typical inputs and assume $M^{}_1$ to lie in the range of 200 GeV to
1 TeV. There are three possible patterns of the light neutrino mass
spectrum: (1) the normal hierarchy: $m^{}_3 \sim 5.0 \times 10^{-2}$
eV, $m^{}_2 \sim 8.8 \times 10^{-3}$ eV, and $m^{}_1$ is much
smaller than $m^{}_2$; (2) the inverted hierarchy: $m^{}_2 \sim 4.9
\times 10^{-2}$ eV, $m^{}_1 \sim 4.8 \times 10^{-2}$ eV, and
$m^{}_3$ is much smaller than $m^{}_1$; (3) the near degeneracy:
$m^{}_1 \sim m^{}_2 \sim m^{}_3 \sim 0.1$ eV to 0.2 eV, which is
consistent with the cosmological upper bound $m^{}_1 + m^{}_2 +
m^{}_3 < 0.67$ eV \cite{WMAP}. In each case, the contributions of
$\nu^{}_i$ and $N^{}_1$ to $(M^{}_{\rm L})^{}_{\alpha\beta}$ in Eq.
(8) will be of the comparable magnitude if the mixing angles
$\theta^{}_{i4}$ satisfy the condition \cite{Ren}
\begin{equation}
s^{}_{i4} s^{}_{j4} \; \sim \; \frac{{\rm max}\{m^{}_1, m^{}_2,
m^{}_3\}}{M^{}_1} \; \sim \; 10^{-14} \cdots 10^{-12} \; ,
\end{equation}
where $i,j = 1,2,3$. This rough estimate allows us to set
$\sqrt{s^{}_{i4} s^{}_{j4}} \sim 10^{-8}$---$10^{-5}$ as the
interference bands of ${\cal B}(H^{+} \to l^+_{\alpha}{\bar
\nu}^{}_{} )$ for $M^{}_1$ to vary between 200 GeV and 1 TeV.
Because the CP-violating phases $\delta^{}_{i4}$ are completely
unrestricted, they may cause either constructive or destructive
effects in the interference bands.

To see the impacts of the Majorana phases on the branching ratios
${\cal B}(H^{+} \to l^+_{\alpha}{\bar \nu}^{}_{})$ in this arresting
parameter region, we may properly redefine the phases of three
charged-lepton fields and then reexpress the neutrino mixing matrix
$V$ in Eq. (7) as
\begin{eqnarray}
V \; = \; \left( \matrix{ c^{}_{14} & 0 & 0 \cr -{s}^{}_{14}
{s}_{24} e^{i\phi} & c^{}_{24} & 0 \cr - {s}^{}_{14} c^{}_{24}
{s}_{34} e^{ i( \phi + \varphi) } & - {s}^{}_{24} {s}_{34}
e^{i\varphi} & c^{}_{34} \cr} \right) V^{}_{0} \; ,
\end{eqnarray}
where
\begin{eqnarray}
V^{}_{0} \; = \; \left( \matrix{ c^{}_{12} c^{}_{13} & {s}^{}_{12}
c^{}_{13} & {s}^{}_{13} e^{-i{\delta}} \cr - {s}^{}_{12} c^{}_{23}
- c^{}_{12} {s}^{}_{13} {s}^{}_{23} e^{i{\delta}} & c^{}_{12}
c^{}_{23} - {s}^{}_{12} {s}^{}_{13} {s}^{}_{23}  e^{i{\delta}} &
c^{}_{13} {s}^{}_{23} \cr {s}^{}_{12} {s}^{}_{23} - c^{}_{12}
{s}^{}_{13} c^{}_{23}  e^{i{\delta}} & -c^{}_{12} {s}^{}_{23} -
{s}^{}_{12} {s}^{}_{13} c^{}_{23}  e^{i{\delta}} & c^{}_{13}
c^{}_{23} } \right) \left( \matrix{ e^{i\rho} & 0 & 0 \cr 0 &
e^{i\sigma} & 0 \cr 0 & 0 & 1 \cr} \right)
\end{eqnarray}
denotes the standard parametrization of the $3\times 3$ unitary
neutrino mixing matrix \cite{PDG}, and the relevant CP-violating
phases are defined as
$\phi={\delta}^{}_{14}-{\delta}^{}_{24}-{\delta}^{}_{12}$,
$\varphi={\delta}^{}_{24}-{\delta}^{}_{34}-{\delta}^{}_{23}$,
$\delta={\delta}^{}_{13}-{\delta}^{}_{12}-{\delta}^{}_{23}$,
$\rho={\delta}^{}_{12}+{\delta}^{}_{23}$ and
$\sigma={\delta}^{}_{23}$. It is clear that $\rho$ and $\sigma$ are
the so-called Majorana phases because they have nothing to do with
neutrino oscillations but may affect the neutrinoless double-beta
decay. With the help of Eqs. (12) and (13), we may rewrite Eqs. (8)
and (9) as follows:
\footnotesize
\begin{eqnarray}
\sum_\beta |\left(M^{}_{\rm L}\right)^{}_{e\beta}|^2 & = & m^2_1
c^2_{12} + m^2_2 s^2_{12} + m^2_3 s^2_{13} + M^2_1
s^2_{14} \left( s^2_{14} + s^2_{24} + s^2_{34} \right) \nonumber \\
&& + 2 m^{}_1 M^{}_1 {\rm Re} \left[ c^{}_{12} {s}^{}_{14}
e^{2i{\delta}^{}_{14}} \left( c^{}_{12} {s}^{}_{14} - {s}^{}_{12}
c^{}_{23} {s}^{}_{24} e^{-i{\phi}^{}_{}} + {s}^{}_{12} {s}^{}_{23}
{s}^{}_{34} e^{-i( \phi + \varphi)} \right) \right] \nonumber \\
&& + 2 m^{}_2 M^{}_1 {\rm Re} \left[ {s}^{}_{12} {s}^{}_{14}
e^{2i({\delta}^{}_{14}- \rho +\sigma)} \left( {s}^{}_{12}
{s}^{}_{14} + c^{}_{12} c^{}_{23} {s}^{}_{24} e^{-i{\phi}^{}_{}} -
c^{}_{12} {s}^{}_{23} {s}^{}_{34} e^{-i( \phi + \varphi) } \right)
\right] \nonumber \\
&& + 2 m^{}_3 M^{}_1 {\rm Re} \left[ {s}^{}_{13} {s}^{}_{14} e^{ i(
2{\delta}^{}_{14} - 2\rho - {\delta}^{}_{} - \phi - \varphi )}
\left( {s}^{}_{13} {s}^{}_{14} e^{i( \phi + \varphi - \delta) } +
{s}^{}_{23} {s}^{}_{24} e^{i{\varphi}^{}_{}} + c^{}_{23} {s}^{}_{34}
\right) \right] \; , \nonumber \\
\sum_\beta |\left(M^{}_{\rm L}\right)^{}_{\mu\beta}|^2 & = & m^2_1
s^2_{12} c^2_{23} + m^2_2 c^2_{12} c^2_{23} + m^2_3 s^2_{23} + M^2_1
s^2_{24} \left( s^2_{14} + s^2_{24} + s^2_{34} \right) \nonumber \\
&& - 2 m^{}_1 M^{}_1 {\rm Re} \left[ {s}^{}_{12} c^{}_{23}
{s}^{}_{24} e^{i(2{\delta}^{}_{14} - \phi^{}_{})} \left( c^{}_{12}
{s}^{}_{14} - {s}^{}_{12} c^{}_{23} {s}^{}_{24} e^{-i{\phi}^{}_{}} +
{s}^{}_{12} {s}^{}_{23} {s}^{}_{34} e^{-i( \phi + \varphi)} \right)
\right] \nonumber \\
&& + 2 m^{}_2 M^{}_1 {\rm Re} \left[ c^{}_{12} c^{}_{23} {s}^{}_{24}
e^{i(2{\delta}^{}_{14}- 2\rho + 2\sigma - \phi^{})} \left(
{s}^{}_{12} {s}^{}_{14} + c^{}_{12} c^{}_{23} {s}^{}_{24}
e^{-i{\phi}^{}_{}} - c^{}_{12} {s}^{}_{23} {s}^{}_{34} e^{-i( \phi +
\varphi)} \right) \right] \nonumber \\
&& + 2 m^{}_3 M^{}_1 {\rm Re} \left[ {s}^{}_{23} {s}^{}_{24} e^{ i (
2{\delta}^{}_{14} - 2\rho - 2\phi - \varphi )} \left( {s}^{}_{13}
{s}^{}_{14} e^{i( \phi + \varphi - \delta) } + {s}^{}_{23}
{s}^{}_{24} e^{i{\varphi}^{}_{}} + c^{}_{23} {s}^{}_{34} \right)
\right] \; , \nonumber \\
\sum_\beta |\left(M^{}_{\rm L}\right)^{}_{\tau\beta}|^2 & = & m^2_1
s^2_{12} s^2_{23} + m^2_2 c^2_{12} s^2_{23} + m^2_3 c^2_{23} + M^2_1
s^2_{34} \left( s^2_{14} + s^2_{24} + s^2_{34} \right) \nonumber \\
&& + 2 m^{}_1 M^{}_1 {\rm Re} \left[ {s}^{}_{12} {s}^{}_{23}
{s}^{}_{34} e^{ i( 2{\delta}^{}_{14} - \phi^{}_{} - \varphi )}
\left( c^{}_{12} {s}^{}_{14} - {s}^{}_{12} c^{}_{23} {s}^{}_{24}
e^{-i{\phi}^{}_{}} + {s}^{}_{12} {s}^{}_{23} {s}^{}_{34} e^{-i( \phi
+ \varphi)} \right)
\right] \nonumber \\
&& - 2 m^{}_2 M^{}_1 {\rm Re} \left[ c^{}_{12} {s}^{}_{23}
{s}^{}_{34} e^{i(2{\delta}^{}_{14} - 2\rho + 2\sigma - \phi^{}
-\varphi)} \left( {s}^{}_{12} {s}^{}_{14} + c^{}_{12} c^{}_{23}
{s}^{}_{24} e^{-i{\phi}^{}_{}} - c^{}_{12} {s}^{}_{23} {s}^{}_{34}
e^{-i( \phi + \varphi)} \right) \right] \nonumber \\
&& + 2 m^{}_3 M^{}_1 {\rm Re} \left[ c^{}_{23} {s}^{}_{34} e^{ 2i (
{\delta}^{}_{14} - \rho - \phi - \varphi )} \left( {s}^{}_{13}
{s}^{}_{14} e^{i( \phi + \varphi - \delta) } + {s}^{}_{23}
{s}^{}_{24} e^{i{\varphi}^{}_{}} + c^{}_{23} {s}^{}_{34} \right)
\right] \; , \nonumber \\
\sum_{\rho,\sigma}|\left(M^{}_{\rm L}\right)^{}_{\rho\sigma}|^2 & =
& m^2_1 + m^2_2 + m^2_3 + M^2_1 \left( s^2_{14} + s^2_{24} +
s^2_{34} \right)^2 \nonumber \\
&& + 2 m^{}_1 M^{}_1 {\rm Re} \left[ e^{i{\delta}^{}_{14}} \left(
c^{}_{12} {s}^{}_{14} - {s}^{}_{12} c^{}_{23} {s}^{}_{24}
e^{-i{\phi}^{}_{}} + {s}^{}_{12} {s}^{}_{23} {s}^{}_{34} e^{-i( \phi
+ \varphi)} \right) \right]^2 \nonumber \\
&& + 2 m^{}_2 M^{}_1 {\rm Re} \left[ e^{i({\delta}^{}_{14} - \rho +
\sigma)} \left( {s}^{}_{12} {s}^{}_{14} + c^{}_{12} c^{}_{23}
{s}^{}_{24} e^{-i{\phi}^{}_{}} - c^{}_{12} {s}^{}_{23} {s}^{}_{34}
e^{-i( \phi + \varphi)} \right) \right]^2 \nonumber \\
&& + 2 m^{}_3 M^{}_1 {\rm Re} \left[ e^{i({\delta}^{}_{14}- \rho)}
\left( {s}^{}_{13} {s}^{}_{14} e^{ -i \delta } + {s}^{}_{23}
{s}^{}_{24} e^{-i{\phi}^{}_{}} + c^{}_{23} {s}^{}_{34} e^{-i( \phi +
\varphi)} \right) \right]^2 \; .
\end{eqnarray}
\normalsize We see that the conventional Majorana phases $\rho$ and
$\sigma$ together with other CP-violating phases show up in the
interference terms. Hence they may affect the branching ratios of
$H^{+} \to l^+_{\alpha}{\bar \nu}^{}$ decays to some extent. We
shall numerically calculate ${\cal B}(H^{+} \to l^+_{\alpha}{\bar
\nu}^{}_{})$ in the subsequent section to illustrate both the
interference bands and the effects of Majorana phases for different
mass spectra of three light neutrinos.

If $M^{}_1 \lesssim {\cal O}(1)$ TeV and the values of $s^{}_{i4}$
lie in the interference bands obtained above, it will be impossible
to produce and observe $N^{}_1$ at the LHC. The reason is simply
that the interaction of $N^{}_1$ with three charged leptons is too
weak to be detected in this parameter space \cite{Ren}. Given the
integrated luminosity to be $100 ~ {\rm fb}^{-1}$, for example, the
resonant signature of $N^{}_1$ in the channel $p\bar{p} \to
\mu^{\pm} N^{}_1$ with $N^{}_1 \to \mu^{\pm}W^{\mp}$ at the LHC has
been analyzed and the sensitivity of the cross section $\sigma
(p\bar{p} \to \mu^{\pm} \mu ^{\pm} W^{\mp}) \approx \sigma(p\bar{p}
\to \mu^\pm N^{}_1) {\cal B}(N^{}_1 \to \mu^{\pm} W^{\mp})$ to the
effective mixing parameter $S^{}_{\mu\mu} \approx s^4_{24}/(s^2_{14}
+ s^2_{24} + s^2_{34})$ has been examined in Ref. \cite{Han}. It is
found that $S^{}_{\mu\mu} \geq 7.2 \times 10^{-4}$ (or equivalently,
$s^2_{24} \geq 2.1 \times 10^{-3}$ for $s^{}_{14} \sim s^{}_{24}
\sim s^{}_{34}$) is required in order to get a signature at the
$2\sigma$ level for $M^{}_1 \geq 200$ GeV. This result illustrates
that there will be no chance to probe the existence of $N^{}_1$ in
the interference bands at the LHC. However, it is possible to
produce $H^{\pm}$ and $H^{\pm\pm}$ at the LHC and to observe the
signatures of $H^{+} \to l^+_{\alpha}{\bar \nu}^{}_{}$, $H^{-} \to
l^-_{\alpha}{\nu}^{}_{}$ and $H^{\pm\pm} \to l^\pm_\alpha
l^\pm_\beta$ decays provided $M^{}_{H^{\pm}} \lesssim {\cal O}(1)$
TeV and $M^{}_{H^{\pm\pm}} \lesssim {\cal O}(1)$ TeV
\cite{Fileviez}. In this case, the measurements of relevant decay
rates or branching ratios are difficult to tell whether the
existence of $H^{\pm}$ and $H^{\pm\pm}$ is due to a pure type-II
seesaw model or due to a (minimal) type-(I+II) seesaw model.

\section{Numerical examples}

For the sake of simplicity, here we take $\theta^{}_{12} = \arctan
(1/\sqrt{2}) \approx 35.3^\circ$, $\theta^{}_{13} = 0^\circ$ and
$\theta^{}_{23} = 45^\circ$; i.e., $V^{}_0$ takes the exact
tri-bimaximal mixing pattern \cite{TB}. The small deviation of $V$
from $V^{}_0$ implies the effect of unitarity violation. We shall do
the numerical calculations in two different ways. Firstly, to
examine the nontrivial role of new CP-violating phases
$\delta^{}_{i4}$ in ${\cal B}(H^{+} \to l^+_{\alpha}{\bar
\nu}^{}_{})$, we switch off the conventional CP-violating phases
$\delta^{}_{12}$, $\delta^{}_{13}$ and $\delta^{}_{23}$. We fix
$\Delta m^2_{21} = 7.7 \times 10^{-5} ~ {\rm eV}^2$, $|\Delta
m^2_{32}| = 2.4 \times 10^{-3} ~ {\rm eV}^2$ and $M^{}_1 = 500$ GeV
in our calculations. To further reduce the number of free
parameters, we shall consider one special case for the mixing angles
$\theta^{}_{i4}$ (e.g., $\theta^{}_{14} = \theta^{}_{24} =
\theta^{}_{34}$) and two special cases for the CP-violating phases
$\delta^{}_{i4}$ (either $\delta^{}_{14} = \delta^{}_{24} =
\delta^{}_{34} =0$ or $\delta^{}_{14} = \delta^{}_{24} =
\delta^{}_{34} =\pi/2$). Secondly, to illustrate the remarkable
effects of two conventional Majorana phases $\rho$ and $\sigma$ on
${\cal B}(H^{+} \to l^+_{\alpha}{\bar \nu}^{}_{})$, we switch off
other CP-violating phases and take $\theta \equiv \theta^{}_{14} =
\theta^{}_{24} = \theta^{}_{34} = 10^{-6.5}$ as a typical input
within the interference bands. Our results and discussions can be
classified into three parts in accordance with three possible mass
patterns of three light neutrinos.

\subsection{Normal hierarchy}

We simply take $m^{}_1 =0$, such that $m^{}_2 \approx 8.8 \times
10^{-3}$ eV and $m^{}_3 \approx 5.0 \times 10^{-2}$ eV can be
extracted from the given values of $\Delta m^2_{21}$ and $|\Delta
m^2_{32}|$. For chosen values of $\theta^{}_{12}$, $\theta^{}_{13}$
and $\theta^{}_{23}$ together with the assumption $\delta^{}_{12} =
\delta^{}_{13} = \delta^{}_{23} = 0$, Eqs. (8) and (9) can now be
simplified to
\begin{eqnarray}
\sum_\beta |\left(M^{}_{\rm L}\right)^{}_{e\beta}|^2 & = &
\frac{1}{3} m^2_2 +  M^{2}_1  s^2_{14} \left( s^2_{14} + s^2_{24} +
s^2_{34} \right) + \frac{2}{3} m^{}_2  M^{}_1  {\rm Re} \left[
\hat{s}^{}_{14} \left( \hat{s}^{}_{14} + \hat{s}^{}_{24} -
\hat{s}^{}_{34} \right) \right] \; ,
\nonumber \\
\sum_\beta |\left(M^{}_{\rm L}\right)^{}_{\mu\beta}|^2 & = &
\frac{1}{3} m^2_2 + \frac{1}{2} m^2_3 + M^{2}_1 s^2_{24} \left(
s^2_{14} + s^2_{24} + s^2_{34} \right) \nonumber
\\ && + \frac{2}{3} m^{}_2  M^{}_1  {\rm Re} \left[ \hat{s}^{}_{24}
\left( \hat{s}^{}_{14} + \hat{s}^{}_{24} - \hat{s}^{}_{34} \right)
\right] + m^{}_3  M^{}_1  {\rm Re} \left[ \hat{s}^{}_{24} \left(
\hat{s}^{}_{24} + \hat{s}^{}_{34} \right) \right] \; ,
\nonumber \\
\sum_\beta |\left(M^{}_{\rm L}\right)^{}_{\tau\beta}|^2 & = &
\frac{1}{3} m^2_2 + \frac{1}{2} m^2_3 + M^{2}_1 s^2_{34} \left(
s^2_{14} + s^2_{24} + s^2_{34} \right) \nonumber
\\ && - \frac{2}{3} m^{}_2  M^{}_1  {\rm Re} \left[ \hat{s}^{}_{34}
\left( \hat{s}^{}_{14} + \hat{s}^{}_{24} - \hat{s}^{}_{34} \right)
\right] + m^{}_3  M^{}_1  {\rm Re} \left[ \hat{s}^{}_{34} \left(
\hat{s}^{}_{24} + \hat{s}^{}_{34} \right) \right] \; ,
\nonumber \\
\sum_{\rho,\sigma}|\left(M^{}_{\rm L}\right)^{}_{\rho\sigma}|^2 & =
& m^2_2 + m^2_3 +  M^{2}_1  \left( s^2_{14} + s^2_{24} + s^2_{34}
\right)^2 \nonumber
\\ && + \frac{2}{3} m^{}_2  M^{}_1 {\rm Re} \left( \hat{s}^{}_{14} +
\hat{s}^{}_{24} - \hat{s}^{}_{34} \right)^2 + m^{}_3 M^{}_1 {\rm Re}
\left( \hat{s}^{}_{24} + \hat{s}^{}_{34} \right)^2 \; .
\end{eqnarray}
Our numerical results for the branching ratios ${\cal B}(H^{+} \to
l^+_{\alpha}{\bar \nu}^{}_{})$ are shown in FIG. 1(a) and FIG. 1(b).

FIG. 1(a) is obtained by taking $\theta^{}_{14} = \theta^{}_{24} =
\theta^{}_{34} \equiv \theta$ and $\delta^{}_{14} = \delta^{}_{24} =
\delta^{}_{34} =0$. We see that ${\cal B}(H^{+} \to {\mu}^+_{}{\bar
\nu}^{}_{}) $ and $ {\cal B}(H^{+} \to {\tau}^+_{}{\bar \nu}^{}_{})$
are approximately the same in the whole parameter space due to an
approximate $\mu$-$\tau$ symmetry.

FIG. 1(b) is obtained by taking $\theta^{}_{14} = \theta^{}_{24} =
\theta^{}_{34} \equiv \theta$ and $\delta^{}_{14} = \delta^{}_{24} =
\delta^{}_{34} =\pi/2$. We see more obvious interference effects for
$\theta$ changing from $10^{-7}$ to $10^{-6}$, which can be
understood with the help of Eqs. (4) and (15). In particular, ${\cal
B}(H^{+} \to {e}^+_{}{\bar \nu}^{}_{})$ is strongly enhanced because
of the destructive interference effect in its denominator, while
${\cal B}(H^{+} \to {\mu}^+_{}{\bar \nu}^{}_{})$ and ${\cal B}(H^{+}
\to {\tau}^+_{}{\bar \nu}^{}_{})$ may reach their minimal values due
to the destructive interference effects in their numerators at
$\theta \sim 2\times 10^{-7}$.

On the other hand, let us simplify Eq. (14) by taking
$\delta^{}_{14} = \phi = \varphi = \delta = 0$:
\begin{eqnarray}
\sum_\beta |\left(M^{}_{\rm L}\right)^{}_{e\beta}|^2  & = &
\frac{1}{3} m^2_2 +  M^{2}_1  s^2_{14} \left( s^2_{14} + s^2_{24} +
s^2_{34} \right) + \frac{2}{3} m^{}_2
 M^{}_1  {s}^{}_{14} \left( {s}^{}_{14} + {s}^{}_{24} -
{s}^{}_{34} \right) \cos{2(\rho-\sigma)} \; ,
\nonumber \\
\sum_\beta |\left(M^{}_{\rm L}\right)^{}_{\mu\beta}|^2 & = &
\frac{1}{3} m^2_2 + \frac{1}{2} m^2_3 +
 M^{2}_1 s^2_{24} \left( s^2_{14} + s^2_{24} + s^2_{34} \right)
 \nonumber \\
 && + \frac{2}{3} m^{}_2 M^{}_1  {s}^{}_{24} \left( {s}^{}_{14} +
{s}^{}_{24} - {s}^{}_{34} \right) \cos{2(\rho-\sigma)} + m^{}_3
M^{}_1  {s}^{}_{24} \left( {s}^{}_{24} + {s}^{}_{34} \right) \cos
{ 2\rho} \; ,
\nonumber \\
\sum_\beta |\left(M^{}_{\rm L}\right)^{}_{\tau\beta}|^2 & = &
\frac{1}{3} m^2_2 + \frac{1}{2} m^2_3 +
 M^{2}_1  s^2_{34} \left( s^2_{14} + s^2_{24} + s^2_{34} \right) \nonumber \\
 && - \frac{2}{3} m^{}_2  M^{}_1  {s}^{}_{34}  \left( {s}^{}_{14} +
{s}^{}_{24} - {s}^{}_{34} \right)  \cos { 2(\rho-\sigma)} + m^{}_3
M^{}_1 {s}^{}_{34} \left( {s}^{}_{24} + {s}^{}_{34} \right) \cos{
2\rho} \; ,
\nonumber \\
\sum_{\rho,\sigma}|\left(M^{}_{\rm L}\right)^{}_{\rho\sigma}|^2 & =
& m^2_2 + m^2_3 +  M^{2}_1  \left( s^2_{14} + s^2_{24} + s^2_{34}
\right)^2 \nonumber \\
&& + \frac{2}{3} m^{}_2  M^{}_1 \left( {s}^{}_{14} + {s}^{}_{24} -
{s}^{}_{34} \right)^2 \cos{ 2(\rho-\sigma)} + m^{}_3 M^{}_1 \left(
{s}^{}_{24} + {s}^{}_{34} \right)^2 \cos { 2\rho} \; .
\end{eqnarray}
Our numerical results for the branching ratios ${\cal B}(H^{+} \to
l^+_{\alpha}{\bar \nu}^{}_{})$ are shown in FIG. 1(c) and FIG. 1(d).

FIG. 1(c) is obtained by taking both $\theta^{}_{14} =
\theta^{}_{24} = \theta^{}_{34} = 10^{-6.5}$ and $\sigma =
\delta^{}_{14}=\phi=\varphi=\delta=0$. We see that ${\cal B}(H^{+}
\to {e}^+_{}{\bar \nu}^{}_{})$, ${\cal B}(H^{+} \to {\mu}^+_{}{\bar
\nu}^{}_{})$ and ${\cal B}(H^{+} \to {\tau}^+_{}{\bar \nu}^{}_{})$
are all sensitive to the Majorana phase $\rho$ changing from $0$ to
$2\pi$.

FIG. 1(d) is obtained by taking $\theta^{}_{14} = \theta^{}_{24} =
\theta^{}_{34} = 10^{-6.5}$ and $\rho =
\delta^{}_{14}=\phi=\varphi=\delta=0$. The slight difference
between ${\cal B}(H^{+} \to {\mu}^+_{}{\bar \nu}^{}_{})$ and
${\cal B}(H^{+} \to {\tau}^+_{}{\bar \nu}^{}_{})$ is easily
understandable from Eq. (16). Compared with FIG. 1(c), FIG. 1(d)
reveals a rather mild dependence of ${\cal B}(H^{+} \to
l^+_{\alpha}{\bar \nu}^{}_{})$ on the Majorana phase $\sigma$. The
reason is simply that the terms proportional to $\cos 2(\rho -
\sigma)$ are more suppressed than those proportional to $\cos
2\rho$ in Eq. (16), as a straightforward result of $m^{}_2 <
m^{}_{3}$.

\subsection{Inverted hierarchy}

We take $m^{}_3 =0$ for simplicity, such that $m^{}_1 \approx 4.8
\times 10^{-2}$ eV and $m^{}_2 \approx 4.9 \times 10^{-2}$ eV can
be extracted from the given values of $\Delta m^2_{21}$ and
$|\Delta m^2_{32}|$. For chosen values of $\theta^{}_{12}$,
$\theta^{}_{13}$ and $\theta^{}_{23}$ together with the assumption
$\delta^{}_{12} = \delta^{}_{13} = \delta^{}_{23} =0$, Eqs. (8)
and (9) can now be simplified to
\begin{eqnarray}
\sum_\beta |\left(M^{}_{\rm L}\right)^{}_{e\beta}|^2  & = &
\frac{2}{3} m^2_1 + \frac{1}{3} m^2_2 + M^{2}_1
s^2_{14} \left( s^2_{14} + s^2_{24} + s^2_{34} \right) \nonumber \\
&& + \frac{2}{3} m^{}_1 M^{}_1  {\rm Re} \left[ \hat{s}^{}_{14}
\left( 2\hat{s}^{}_{14} - \hat{s}^{}_{24} + \hat{s}^{}_{34} \right)
\right] + \frac{2}{3} m^{}_2  M^{}_1  {\rm Re} \left[ \hat{s}^{}_{14}
\left( \hat{s}^{}_{14} + \hat{s}^{}_{24} - \hat{s}^{}_{34} \right)
\right] \; ,
\nonumber \\
\sum_\beta |\left(M^{}_{\rm L}\right)^{}_{\mu\beta}|^2 & = &
\frac{1}{6} m^2_1 + \frac{1}{3} m^2_2 + M^{2}_1
s^2_{24} \left( s^2_{14} + s^2_{24} + s^2_{34} \right) \nonumber \\
&& - \frac{1}{3} m^{}_1  M^{}_1  {\rm Re} \left[ \hat{s}^{}_{24}
\left( 2 \hat{s}^{}_{14} - \hat{s}^{}_{24} + \hat{s}^{}_{34} \right)
\right] + \frac{2}{3} m^{}_2 M^{}_1  {\rm Re} \left[ \hat{s}^{}_{24}
\left( \hat{s}^{}_{14} + \hat{s}^{}_{24} - \hat{s}^{}_{34} \right)
\right] \; ,
\nonumber \\
\sum_\beta |\left(M^{}_{\rm L}\right)^{}_{\tau\beta}|^2  & = &
\frac{1}{6} m^2_1 + \frac{1}{3} m^2_2 + M^{2}_1
s^2_{34} \left( s^2_{14} + s^2_{24} + s^2_{34} \right) \nonumber \\
&& + \frac{1}{3} m^{}_1  M^{}_1  {\rm Re} \left[ \hat{s}^{}_{34}
\left( 2 \hat{s}^{}_{14} - \hat{s}^{}_{24} + \hat{s}^{}_{34} \right)
\right] - \frac{2}{3} m^{}_2  M^{}_1  {\rm Re} \left[ \hat{s}^{}_{34}
\left( \hat{s}^{}_{14} + \hat{s}^{}_{24} - \hat{s}^{}_{34} \right)
\right] \; ,
\nonumber \\
\sum_{\rho,\sigma}|\left(M^{}_{\rm L}\right)^{}_{\rho\sigma}|^2 &
= & m^2_1 + m^2_2 +  M^{2}_1
\left( s^2_{14} + s^2_{24} + s^2_{34} \right)^2 \nonumber \\
&& + \frac{1}{3} m^{}_1  M^{}_1  {\rm Re}  \left(
2\hat{s}^{}_{14} - \hat{s}^{}_{24} + \hat{s}^{}_{34} \right)^2 +
\frac{2}{3} m^{}_2  M^{}_1  {\rm Re}  \left( \hat{s}^{}_{14} +
\hat{s}^{}_{24} - \hat{s}^{}_{34} \right)^2 \; .
\end{eqnarray}
Our numerical results for the branching ratios ${\cal B}(H^{+} \to
l^+_{\alpha}{\bar \nu}^{}_{})$ are shown in FIG. 2(a) and FIG.
2(b).

FIG. 2(a) is obtained by taking $\theta^{}_{14} = \theta^{}_{24} =
\theta^{}_{34} \equiv \theta$ and $\delta^{}_{14} = \delta^{}_{24}
= \delta^{}_{34} =0$. We see that ${\cal B}(H^{+} \to
{\mu}^+_{}{\bar \nu}^{}_{}) = {\cal B}(H^{+} \to {\tau}^+_{}{\bar
\nu}^{}_{})$ holds in the whole parameter space due to
$\mu$-$\tau$ symmetry.

FIG. 2(b) is obtained by taking $\theta^{}_{14} = \theta^{}_{24} =
\theta^{}_{34} \equiv \theta$ and $\delta^{}_{14} = \delta^{}_{24}
= \delta^{}_{34} =\pi/2$. One can see more obvious interference
effects for $\theta$ changing from $10^{-7}$ to $10^{-6}$, which
can be understood with the help of Eqs. (4) and (17). In
particular, ${\cal B}(H^{+} \to {e}^+_{}{\bar \nu}^{}_{})$
undergoes a minimum because of the destructive interference effect
in its numerator, while ${\cal B}(H^{+} \to {\mu}^+_{}{\bar
\nu}^{}_{})$ or ${\cal B}(H^{+} \to {\tau}^+_{}{\bar \nu}^{}_{})$
undergoes a maximum due to the destructive interference effect in
its denominator when $\theta$ varies in the interference band.

On the other hand, we simplify Eq. (14) by taking $\delta^{}_{14}
= \phi = \varphi = \delta = 0$:
\begin{eqnarray}
\sum_\beta |\left(M^{}_{\rm L}\right)^{}_{e\beta}|^2 & = &
\frac{2}{3} m^2_1 + \frac{1}{3} m^2_2 +
 M^{2}_1  s^2_{14} \left( s^2_{14} + s^2_{24} + s^2_{34} \right) +
\frac{2}{3} m^{}_1  M^{}_1 {s}^{}_{14} \left( 2{s}^{}_{14}
- {s}^{}_{24} + {s}^{}_{34} \right) \nonumber \\
&& + \frac{2}{3} m^{}_2 M^{}_1 {s}^{}_{14} \left( {s}^{}_{14} +
{s}^{}_{24} - {s}^{}_{34} \right) \cos{2(\rho-\sigma)} \; ,
\nonumber \\
\sum_\beta |\left(M^{}_{\rm L}\right)^{}_{\mu\beta}|^2 & = &
\frac{1}{6} m^2_1 + \frac{1}{3} m^2_2 +
 M^{2}_1  s^2_{24} \left( s^2_{14} + s^2_{24} + s^2_{34} \right) -
\frac{1}{3} m^{}_1 M^{}_1  {s}^{}_{24} \left( 2
{s}^{}_{14} - {s}^{}_{24} + {s}^{}_{34} \right) \nonumber \\
&& + \frac{2}{3} m^{}_2  M^{}_1  {s}^{}_{24} \left( {s}^{}_{14} +
{s}^{}_{24} - {s}^{}_{34} \right) \cos{2(\rho-\sigma)} \; ,
\nonumber \\
\sum_\beta |\left(M^{}_{\rm L}\right)^{}_{\tau\beta}|^2 & = &
\frac{1}{6} m^2_1 + \frac{1}{3} m^2_2 +
 M^{2}_1  s^2_{34} \left( s^2_{14} + s^2_{24} + s^2_{34} \right) +
\frac{1}{3} m^{}_1  M^{}_1  {s}^{}_{34} \left( 2
{s}^{}_{14} - {s}^{}_{24} + {s}^{}_{34} \right) \nonumber \\
&& - \frac{2}{3} m^{}_2  M^{}_1  {s}^{}_{34} \left( {s}^{}_{14} +
{s}^{}_{24} - {s}^{}_{34} \right) \cos{2(\rho-\sigma)} \; ,
\nonumber \\
\sum_{\rho,\sigma}|\left(M^{}_{\rm L}\right)^{}_{\rho\sigma}|^2 &
= & m^2_1 + m^2_2 +  M^{2}_1  \left( s^2_{14} + s^2_{24} +
s^2_{34} \right)^2 + \frac{1}{3} m^{}_1  M^{}_1
\left( 2{s}^{}_{14} - {s}^{}_{24} + {s}^{}_{34} \right)^2 \nonumber \\
&& + \frac{2}{3} m^{}_2  M^{}_1  \left( {s}^{}_{14} + {s}^{}_{24}
- {s}^{}_{34} \right)^2 \cos{2(\rho-\sigma)} \; .
\end{eqnarray}
Our numerical results for the branching ratios ${\cal B}(H^{+} \to
l^+_{\alpha}{\bar \nu}^{}_{})$ are shown in FIG. 2(c) and FIG.
2(d).

FIG. 2(c) is obtained by taking $\theta^{}_{14} = \theta^{}_{24} =
\theta^{}_{34} = 10^{-6.5}$ and $\sigma = \delta^{}_{14} = \phi =
\varphi = \delta = 0$. We see that ${\cal B}(H^{+} \to
{e}^+_{}{\bar \nu}^{}_{})$, ${\cal B}(H^{+} \to {\mu}^+_{}{\bar
\nu}^{}_{})$ and ${\cal B}(H^{+} \to {\tau}^+_{}{\bar \nu}^{}_{})$
are all sensitive to the Majorana phase $\rho$ varying from $0$ to
$2\pi$. FIG. 2(d) is obtained by taking $\theta^{}_{14} =
\theta^{}_{24} = \theta^{}_{34} = 10^{-6.5}$ and $\rho =
\delta^{}_{14} = \phi = \varphi = \delta = 0$. Hence the results
of ${\cal B}(H^{+} \to l^+_{\alpha}{\bar \nu}^{}_{})$ in FIG. 2(d)
are the same as those in FIG. 2(c), as a straightforward
consequence of the $\rho$-$\sigma$ permutation symmetry which can
be seen from Eq. (18).

\subsection{Near degeneracy}

We assume $m^{}_{1}\approx m^{}_{2}\approx m^{}_{3}\approx 0.1 ~
{\rm eV}$. Then $m^{}_2 - m^{}_1 \approx 3.9 \times 10^{-4}$ eV
and $m^{}_3 - m^{}_2 \approx \pm 1.2 \times 10^{-2}$ eV can be
extracted from given values of $\Delta m^2_{21}$ and $|\Delta
m^2_{32}|$, respectively. For chosen values of $\theta^{}_{12}$,
$\theta^{}_{13}$ and $\theta^{}_{23}$ together with the assumption
$\delta^{}_{12} = \delta^{}_{13} = \delta^{}_{23} =0$, Eqs. (8)
and (9) can now be simplified to
\begin{eqnarray}
\sum_\beta |\left(M^{}_{\rm L}\right)^{}_{e\beta}|^2 & \approx &
m^2_1 +  M^{2}_1  s^2_{14} \left( s^2_{14} + s^2_{24} + s^2_{34}
\right) + 2 m^{}_1  M^{}_1 {s}^{2}_{14} ~ \cos 2\delta^{}_{14} \;
,
\nonumber \\
\sum_\beta |\left(M^{}_{\rm L}\right)^{}_{\mu\beta}|^2 & \approx &
m^2_1 +  M^{2}_1  s^2_{24} \left( s^2_{14} + s^2_{24} + s^2_{34}
\right) + 2 m^{}_1  M^{}_1 {s}^{2}_{24} ~ \cos 2\delta^{}_{24} \;
,
\nonumber \\
\sum_\beta |\left(M^{}_{\rm L}\right)^{}_{\tau\beta}|^2 & \approx
& m^2_1 +  M^{2}_1  s^2_{34} \left( s^2_{14} + s^2_{24} + s^2_{34}
\right) + 2 m^{}_1  M^{}_1 {s}^{2}_{34} ~ \cos 2\delta^{}_{34} \;
,
\nonumber \\
\sum_{\rho,\sigma}|\left(M^{}_{\rm L}\right)^{}_{\rho\sigma}|^2 &
\approx & 3m^2_1 +  M^{2}_1  \left( s^2_{14} + s^2_{24} + s^2_{34}
\right)^2 \nonumber \\
&& + 2 m^{}_1  M^{}_1  \left( s^2_{14} ~ \cos 2\delta^{}_{14} +
s^2_{24} ~ \cos 2\delta^{}_{24} + s^2_{34} ~ \cos 2\delta^{}_{34}
\right) \; ,
\end{eqnarray}
where we have omitted the small mass differences of $\nu^{}_i$. We
fix $m^{}_3 > m^{}_2$ and keep two small mass differences in our
numerical calculations. The results for ${\cal B}(H^{+} \to
l^+_{\alpha}{\bar \nu}^{}_{})$ are shown in FIG. 3(a) and FIG.
3(b).

FIG. 3(a) is obtained by taking $\theta^{}_{14} = \theta^{}_{24} =
\theta^{}_{34} \equiv \theta$ and $\delta^{}_{14} = \delta^{}_{24}
= \delta^{}_{34} =0$. We find that ${\cal B}(H^{+} \to
{e}^+_{}{\bar \nu}^{}_{}) \approx {\cal B}(H^{+} \to
{\mu}^+_{}{\bar \nu}^{}_{}) \approx {\cal B}(H^{+} \to
{\tau}^+_{}{\bar \nu}^{}_{})$ approximately holds in the whole
parameter space, as one can simply see from Eq. (19). Similar
results are also obtained in FIG. 3(b), where $\theta^{}_{14} =
\theta^{}_{24} = \theta^{}_{34} \equiv \theta$ and $\delta^{}_{14}
= \delta^{}_{24} = \delta^{}_{34} =\pi/2$ have been taken. In both
cases, the changes of ${\cal B}(H^{+} \to l^+_{\alpha}{\bar
\nu}^{}_{})$ with $\theta$ are very mild.

On the other hand, we simplify Eq. (14) by taking $\delta^{}_{14}
= \phi = \varphi = \delta = 0$:
\begin{eqnarray}
\sum_\beta |\left(M^{}_{\rm L}\right)^{}_{e\beta}|^2  & \approx &
m^2_1 +  M^{2}_1  s^2_{14} \left( s^2_{14} + s^2_{24} + s^2_{34}
\right) + \frac{2}{3} m^{}_1 M^{}_1  {s}^{}_{14} \left[ \left(
2{s}^{}_{14} - {s}^{}_{24} + {s}^{}_{34} \right) \right. \nonumber
\\
&& \left. + \left( {s}^{}_{14} + {s}^{}_{24} - {s}^{}_{34} \right)
{\rm cos}{2(\rho-\sigma)} \right] \; ,
\nonumber \\
\sum_\beta |\left(M^{}_{\rm L}\right)^{}_{\mu\beta}|^2 & \approx &
m^2_1 +  M^{2}_1  s^2_{24} \left( s^2_{14} + s^2_{24} + s^2_{34}
\right) - \frac{1}{3} m^{}_1 M^{}_1 {s}^{}_{24} \left[ \left(
2{s}^{}_{14} - {s}^{}_{24} + {s}^{}_{34} \right) \right. \nonumber
\\
&& \left. - 2 \left( {s}^{}_{14} + {s}^{}_{24} - {s}^{}_{34}
\right) {\rm cos}{2(\rho-\sigma)} - 3 \left( {s}^{}_{24} +
{s}^{}_{34} \right) {\rm cos}{ 2\rho} \right] \; ,
\nonumber \\
\sum_\beta |\left(M^{}_{\rm L}\right)^{}_{\tau\beta}|^2 & \approx
& m^2_1 +  M^{2}_1  s^2_{34} \left( s^2_{14} + s^2_{24} + s^2_{34}
\right) + \frac{1}{3} m^{}_1  M^{}_1  {s}^{}_{34} \left[ \left(
2{s}^{}_{14} - {s}^{}_{24} + {s}^{}_{34} \right) \right. \nonumber
\\
&& \left. - 2 \left( {s}^{}_{14} + {s}^{}_{24} - {s}^{}_{34}
\right) {\rm cos}{2(\rho-\sigma)} + 3 \left( {s}^{}_{24} +
{s}^{}_{34} \right) {\rm cos}{ 2\rho} \right] \; ,
\nonumber \\
\sum_{\rho,\sigma}|\left(M^{}_{\rm L}\right)^{}_{\rho\sigma}|^2 &
\approx & 3m^2_1 +  M^{2}_1  \left( s^2_{14} + s^2_{24} + s^2_{34}
\right)^2 + \frac{1}{3} m^{}_1  M^{}_1 \left[ \left( 2 {s}^{}_{14}
- {s}^{}_{24} + {s}^{}_{34} \right)^2 \right. \nonumber \\
&& \left. + 2 \left( {s}^{}_{14} + {s}^{}_{24} - {s}^{}_{34}
\right)^2 {\rm cos}{2(\rho-\sigma)} + 3 \left( {s}^{}_{24} +
{s}^{}_{34} \right)^2 {\rm cos}{ 2\rho} \right] \; ,
\end{eqnarray}
where we have omitted the small mass differences of $\nu^{}_i$. We
fix $m^{}_3 > m^{}_2$ and keep two small mass differences in our
numerical calculations. The results for ${\cal B}(H^{+} \to
l^+_{\alpha}{\bar \nu}^{}_{})$ are shown in FIG. 3(c) and FIG.
3(d).

FIG. 3(c) is obtained by taking $\theta^{}_{14} = \theta^{}_{24} =
\theta^{}_{34} = 10^{-6.5}$ and $\sigma =
\delta^{}_{14}=\phi=\varphi=\delta=0$. We see that ${\cal B}(H^{+}
\to {e}^+_{}{\bar \nu}^{}_{})$, ${\cal B}(H^{+} \to
{\mu}^+_{}{\bar \nu}^{}_{})$ and ${\cal B}(H^{+} \to
{\tau}^+_{}{\bar \nu}^{}_{})$ are all sensitive to the Majorana
phase $\rho$ changing from $0$ to $2\pi$.

FIG. 3(d) is obtained by taking $\theta^{}_{14} = \theta^{}_{24} =
\theta^{}_{34} = 10^{-6.5}$ and $\rho =
\delta^{}_{14}=\phi=\varphi=\delta=0$. We see that the behaviors
of ${\cal B}(H^{+} \to {e}^+_{}{\bar \nu}^{}_{})$, ${\cal B}(H^{+}
\to {\mu}^+_{}{\bar \nu}^{}_{})$ and ${\cal B}(H^{+} \to
{\tau}^+_{}{\bar \nu}^{}_{})$ changing with the Majorana phase
$\sigma$ are different from and milder than those in FIG. 3(c), as
one can easily understand from Eq. (20).

\section{Summary}

We have studied the lepton-number-violating decays of
singly-charged Higgs bosons $H^{\pm}$ in the minimal type-(I+II)
seesaw model with one heavy Majorana neutrino $N^{}_1$ and one
$SU(2)_L$ Higgs triplet $\Delta$ at the TeV scale. Their branching
ratios ${\cal B}(H^{+} \to l^+_{\alpha}{\bar \nu}^{}_{})$ depend
not only on the masses, flavor mixing angles and CP-violating
phases of three light neutrinos $\nu^{}_i$ (for $i=1,2,3$) but
also on those of $N^{}_1$. We have focused our attention on the
interference bands of ${\cal B}(H^{+} \to l^+_{\alpha}{\bar
\nu}^{}_{})$, in which the contributions of light and heavy
Majorana neutrinos are comparable in magnitude. We emphasize that
both constructive and destructive interference effects are
possible in the interference bands, and thus it is very difficult
to distinguish the (minimal) type-(I+II) seesaw model from the
type-II seesaw model in this parameter space. While the
lepton-number-violating decays of $H^\pm$ are independent of the
conventional Majorana phases $\rho$ and $\sigma$ in the type-II
seesaw mechanism, they {\it do} depend on $\rho$ and $\sigma$ in
the type-(I+II) seesaw scenario. Although our numerical results
are subject to a simplified type-(I+II) seesaw model, they can
serve as a good example to illustrate the interplay between type-I
and type-II seesaw terms in a generic type-(I+II) seesaw framework
which involves more free parameters.

\vspace{0.5cm}

This work was supported in part by the National Natural Science
Foundation of China. We are grateful to W. Chao and S. Zhou for
useful discussions.

\newpage

\begin{figure}
\psfig{file=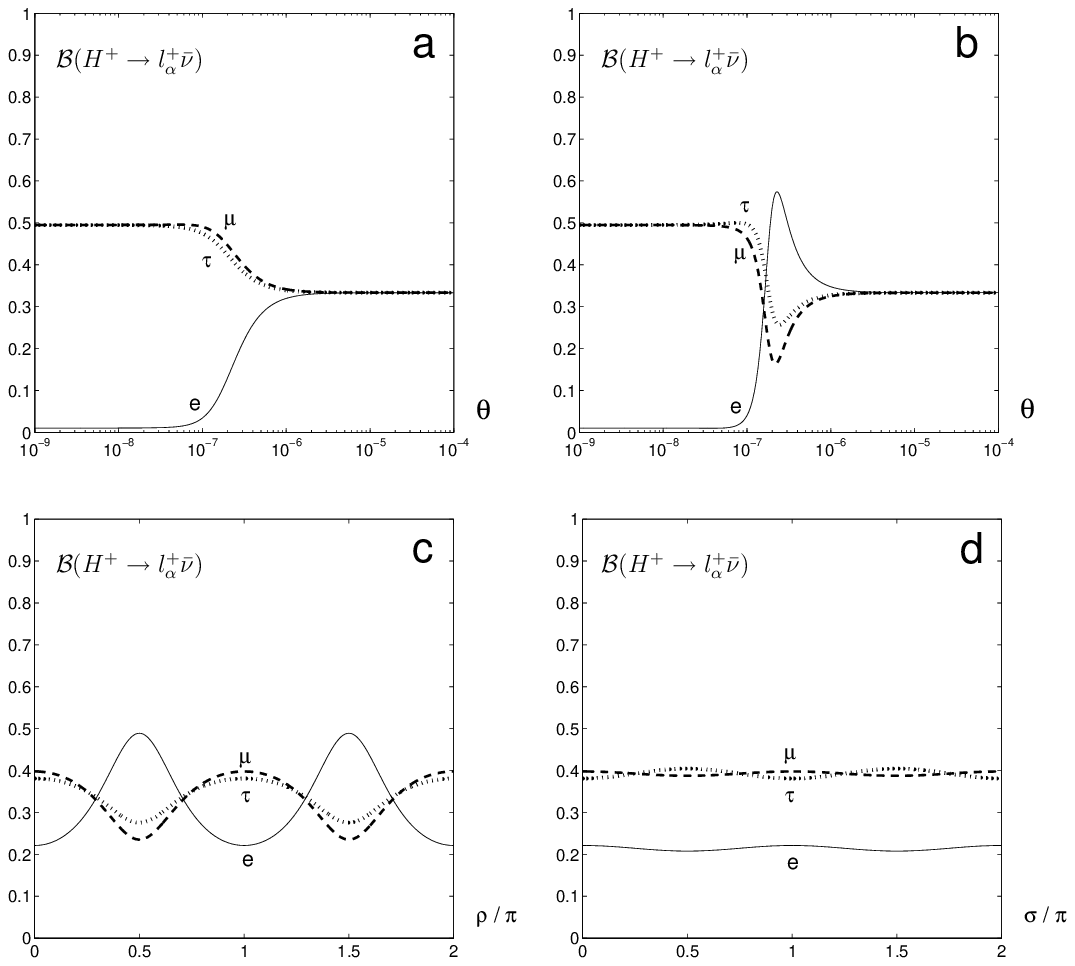, bbllx=2.5cm, bblly=19.6cm, bburx=6.6cm,
bbury=24.0cm, width=3.8cm, height=4cm, angle=0, clip=0}
\vspace{13.5cm} \caption{Branching ratios of $H^{+} \to
l^+_{\alpha}{\bar \nu}^{}_{}$ decays for the normal hierarchy of
$m^{}_i$ with $m^{}_1 =0$: (a) $\theta^{}_{14} = \theta^{}_{24} =
\theta^{}_{34} \equiv \theta$ and $\delta^{}_{14} = \delta^{}_{24} =
\delta^{}_{34} =0$; (b) $\theta^{}_{14} = \theta^{}_{24} =
\theta^{}_{34} \equiv \theta$ and $\delta^{}_{14} = \delta^{}_{24} =
\delta^{}_{34} =\pi/2$;  (c) $\theta^{}_{14} = \theta^{}_{24} =
\theta^{}_{34} = 10^{-6.5}$ and
$\delta^{}_{14}=\phi=\varphi=\delta=\sigma =  0$; (d)
$\theta^{}_{14} = \theta^{}_{24} = \theta^{}_{34} = 10^{-6.5}$ and
$\delta^{}_{14}=\phi=\varphi=\delta=\rho =  0$. }
\end{figure}

\begin{figure}
\psfig{file=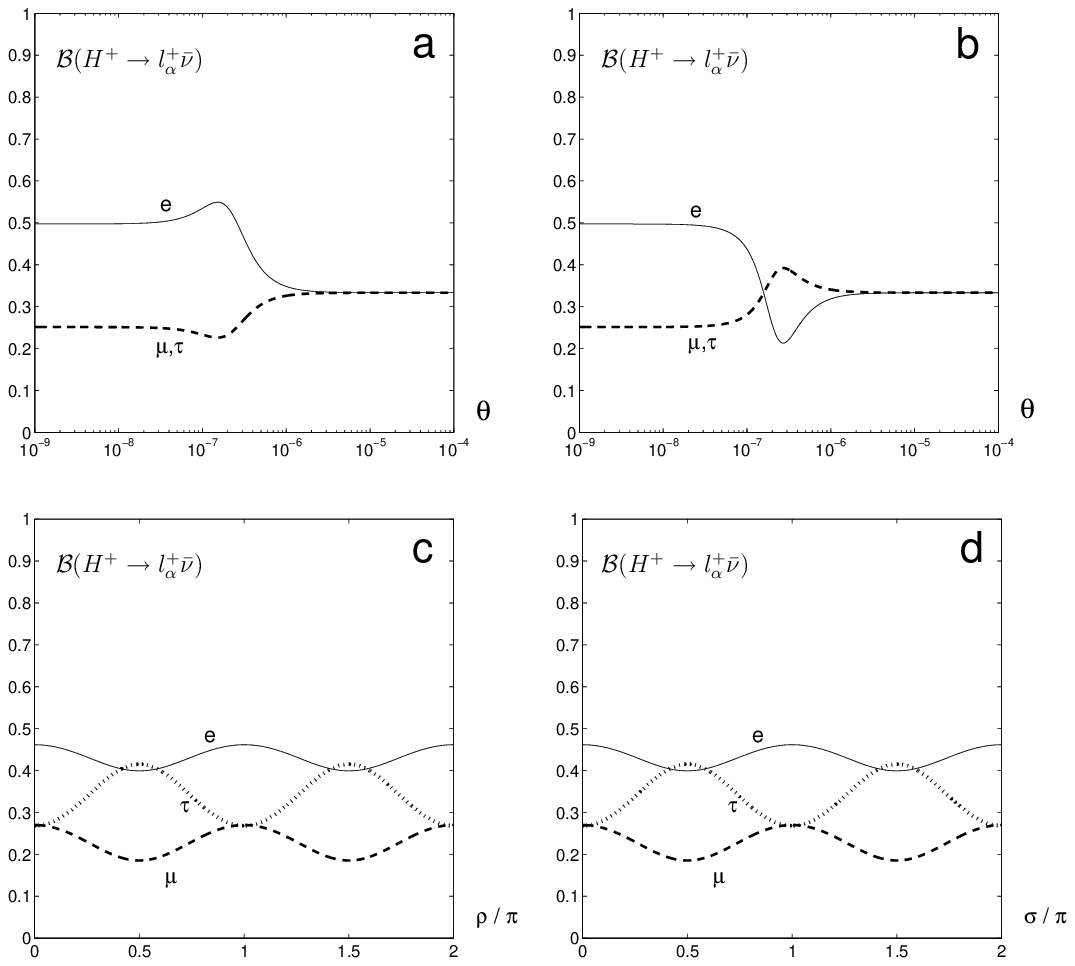, bbllx=2.5cm, bblly=19.6cm, bburx=6.6cm,
bbury=24.0cm, width=3.8cm, height=4cm, angle=0, clip=0}
\vspace{13.5cm} \caption{Branching ratios of $H^{+} \to
l^+_{\alpha}{\bar \nu}^{}_{}$ decays for the inverted hierarchy of
$m^{}_i$ with $m^{}_3 =0$:   (a) $\theta^{}_{14} = \theta^{}_{24} =
\theta^{}_{34} \equiv \theta$ and $\delta^{}_{14} = \delta^{}_{24} =
\delta^{}_{34} =0$;   (b) $\theta^{}_{14} = \theta^{}_{24} =
\theta^{}_{34} \equiv \theta$ and $\delta^{}_{14} = \delta^{}_{24} =
\delta^{}_{34} =\pi/2$;  (c) $\theta^{}_{14} = \theta^{}_{24} =
\theta^{}_{34}  = 10^{-6.5}$ and
$\delta^{}_{14}=\phi=\varphi=\delta=\sigma =  0$; (d)
$\theta^{}_{14} = \theta^{}_{24} = \theta^{}_{34}  = 10^{-6.5}$ and
$\delta^{}_{14}=\phi=\varphi=\delta=\rho =  0$. }
\end{figure}

\begin{figure}
\psfig{file=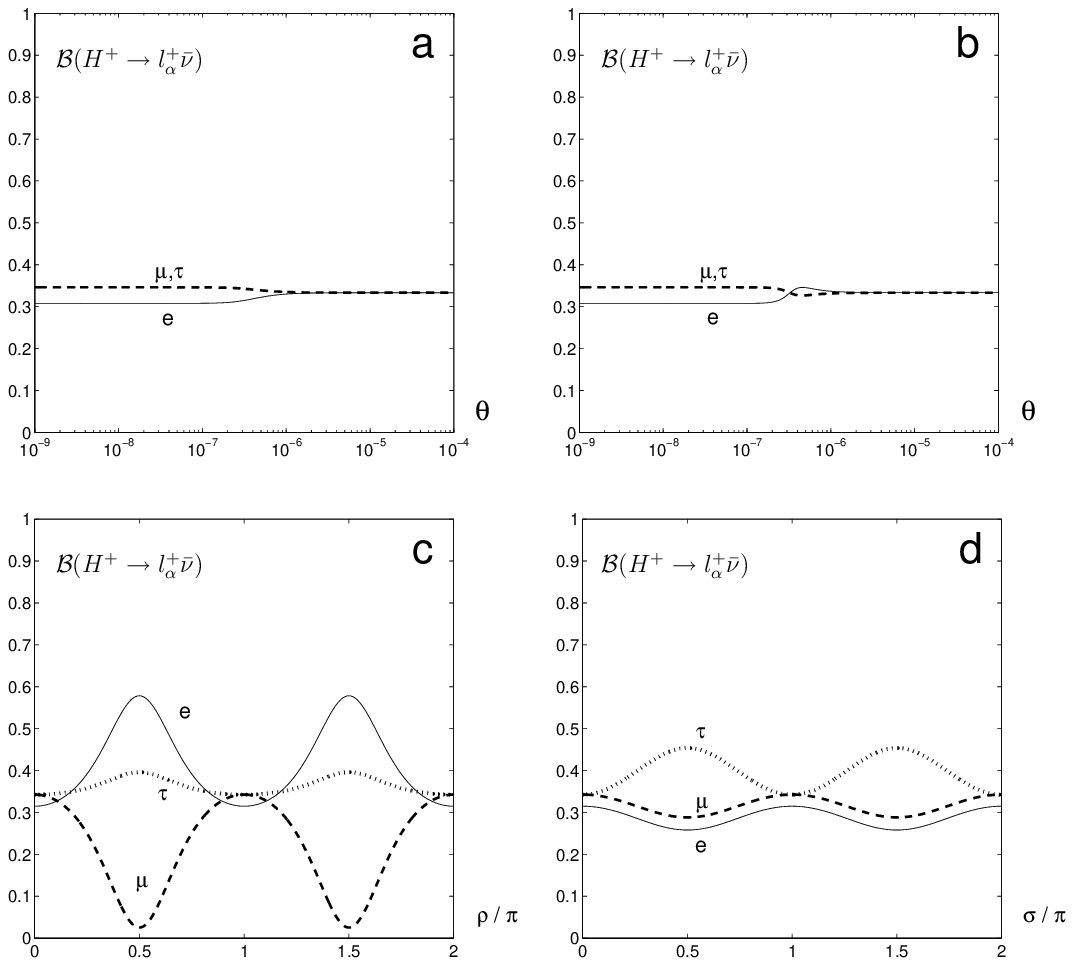, bbllx=2.5cm, bblly=19.6cm, bburx=6.6cm,
bbury=24.0cm, width=3.8cm, height=4cm, angle=0, clip=0}
\vspace{13.5cm} \caption{Branching ratios of $H^{+} \to
l^+_{\alpha}{\bar \nu}^{}_{}$ decays for the near degeneracy of
$m^{}_i$ with $m^{}_3 > m^{}_2$:   (a) $\theta^{}_{14} =
\theta^{}_{24} = \theta^{}_{34} \equiv \theta$ and $\delta^{}_{14} =
\delta^{}_{24} = \delta^{}_{34} =0$; (b) $\theta^{}_{14} =
\theta^{}_{24} = \theta^{}_{34} \equiv \theta$ and $\delta^{}_{14} =
\delta^{}_{24} = \delta^{}_{34} =\pi/2$;  (c) $\theta^{}_{14} =
\theta^{}_{24} = \theta^{}_{34}  = 10^{-6.5}$ and
$\delta^{}_{14}=\phi=\varphi=\delta=\sigma =  0$; (d)
$\theta^{}_{14} = \theta^{}_{24} = \theta^{}_{34}  = 10^{-6.5}$ and
$\delta^{}_{14}=\phi=\varphi=\delta=\rho =  0$. }
\end{figure}

\end{document}